\documentclass{Interspeech}
\usepackage{multirow}
\interspeechcameraready 
\title{Enhancing Speech Emotion Recognition with Multi-Task Learning and Dynamic Feature Fusion}

\author{Honghong}{Wang}
\author{Jing}{Deng}
\author{Fanqin}{Meng}
\author{Rong}{Zheng}

\affiliation[nocounter]{}{Beijing Fosafer Information Technology Co., Ltd.}{China}
\email{\{wanghonghong, dengjing, mengfanqin, zhengrong\}@fosafer.com}

\usepackage{comment}

\begin{document}

\maketitle

\begin{abstract}
    
    This study investigates fine-tuning self-supervised learning (SSL) models using multi-task learning (MTL) to enhance speech emotion recognition (SER). The framework simultaneously handles four related tasks: emotion recognition, gender recognition, speaker verification, and automatic speech recognition. An innovative co-attention module is introduced to dynamically capture the interactions between features from the primary emotion classification task and auxiliary tasks, enabling context-aware fusion. Moreover, We introduce the Sample Weighted Focal Contrastive (SWFC) loss function to address class imbalance and semantic confusion by adjusting sample weights for difficult and minority samples. The method is validated on the Categorical Emotion Recognition task of the Speech Emotion Recognition in Naturalistic Conditions Challenge, showing significant performance improvements. 
\end{abstract}

\section{Introduction}

Recently, the rapid advancement of artificial intelligence has significantly improved the role of speech in human-computer interaction (HCI) \cite{b1}. As a crucial element of HCI, Speech Emotion Recognition (SER) has emerged as a key area of research in multimodal emotion analysis, with the goal of accurately identifying and interpreting the speaker's emotional state.

SER currently encounters several challenges. A major challenge is the scarcity of high-quality emotion-labeled datasets. Furthermore, the emotional features of speech are dynamic and varying over time, with valuable information distributed unevenly in the domain of time and frequency \cite{b2}. Traditional deep learning models, such as convolutional neural networks (CNNs) \cite{b3}, are constrained by a fixed perceptual field, making it difficult to accurately capture emotional cues, which often leads to subpar recognition performance. To address these challenges, researchers have increasingly turned to self-supervised learning (SSL) models, leading to the development of powerful end-to-end speech representation models like wav2vec 2.0 \cite{b4}, WavLM \cite{b5}, and HuBERT \cite{b6}. These models improve emotion recognition by extracting high-dimensional representations from raw waveform signals, leveraging large-scale unsupervised pre-training to capture rich semantic, phonetic, and paralinguistic information \cite{b7}.

Since paralinguistic and acoustic features are believed to carry personalized emotional information \cite{b8}, researchers have increasingly turned to multi-task learning (MTL) approaches \cite{b9,b10}, which simultaneously train the SER task alongside these auxiliary tasks. By sharing parameters across multiple tasks, the model can better extract emotional information from speech features, resulting in significant improvements in SER performance \cite{b11}. However, traditional MTL methods often fail to fully take advantage of the representations in each auxiliary task. As a result, integrating information from paralinguistic factors, such as speaker identity, gender, and speech content, with the emotion branch has been shown to further enhance emotion recognition performance \cite{b12}.

The objective of the Speech Emotion Recognition in Naturalistic Conditions Challenge \cite{b13} is to predict the emotional state of speakers in real-life scenarios. This challenge consists of two tracks: (1) Categorical Emotion Recognition and (2) Emotional Attributes Prediction. The Categorical Emotion Recognition track involves classifying each sample into one of eight emotional categories: anger, happiness, sadness, fear, surprise, contempt, disgust, and neutral. As shown in Figure 1, the training and validation data for the challenge are typically imbalanced.

\begin{figure}[htbp]
\centerline{\includegraphics[width=0.35\textwidth]{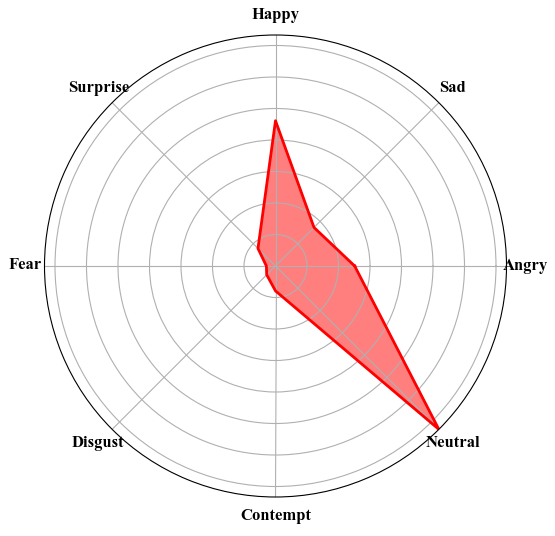}} % 调整图片宽度为页面宽度
\caption{Distribution of the challenge training and validation sets  across the eight emotional labels.}
\label{fig}
\end{figure}

This paper presents a SER model that employs a MTL approach. The model incorporates a co-attention module to integrate emotion, gender, speaker identity, and speech content, enabling it to capture the relationships between emotional information and other auxiliary factors. To address the challenges of recognizing a limited number of categories and mitigating semantic confusion caused by the long-tailed distribution of emotion categories in the challenge data, we introduce the Sample Weighted Focal Contrastive (SWFC) loss function \cite{b15}. We have performed a comprehensive evaluation and demonstrated the effectiveness of the proposed approach.

\section{Proposed Method}

\subsection{Model Architecture}

As shown in Figure 2, our research framework consists of three key stages: feature extraction, pooling, and classification. The network follows an end-to-end structure, where raw speech is first up-sampled using a convolutional feature encoder. Then, deep feature extraction is performed by a Transformer Encoder. Recognizing the varying importance of different feature layers in the SSL model for different downstream tasks, we employ a learnable weighting sum method \cite{b16}. This method adaptively integrates the hidden representations from all layers of the SSL model at the utterance level to generate a unified representation.

\begin{figure}[htbp]
\centerline{\includegraphics[width=0.45\textwidth]{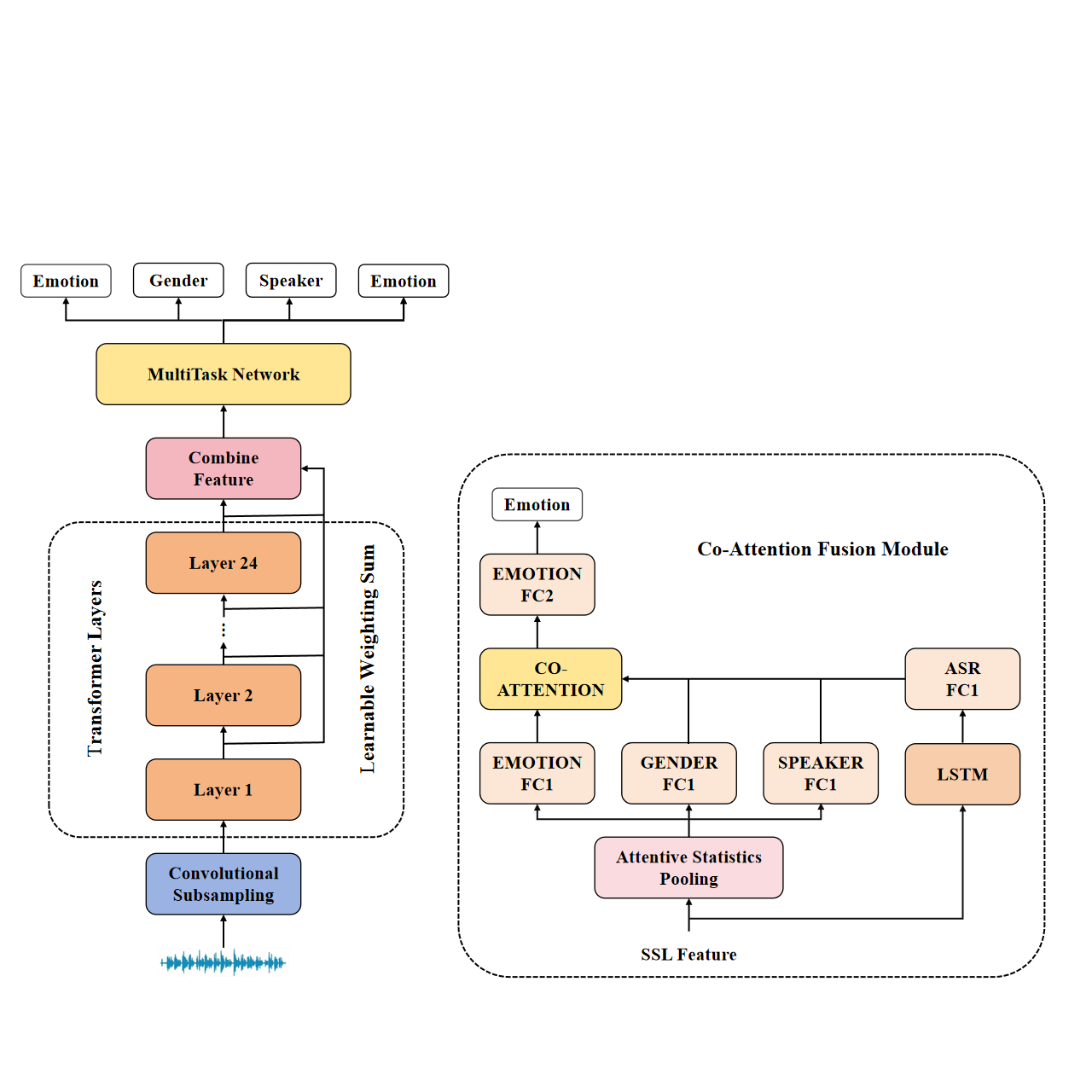}} % 调整图片宽度为页面宽度
\caption{Network structure of our proposed MTL method.}
\label{fig}
\end{figure}

The emotion, gender and speaker recognition branches take the global representations encoded by the attentive statistics pooling layer \cite{b16} as input, while the automatic speech recognition (ASR) branch uses the high-dimensional representations of the SSL model directly. The multi-task branching network then predicts emotion, gender, speaker, and ASR outputs, respectively. Notably, the emotion recognition branch incorporates a Co-attention module that facilitates dynamic interaction between emotion features and other auxiliary task features.

During training, the emotion, gender, and speaker branches are optimized using a cross-entropy loss function, while the ASR branch uses a connectionist temporal classification (CTC) loss function \cite{b17}. To further enhance performance, we introduce a SWFC loss function, which operates on the global representation output from the SSL model. This loss function enables the SSL model to focus more on challenging, less frequent samples during fine-tuning. The overall objective function L of the model is defined as:
\begin{align}
L &= (1 - 3\alpha) L_{\mathrm{emotion}} + \alpha L_{\mathrm{gender}} \nonumber \\
  &\quad + \alpha L_{\mathrm{speaker}} + \alpha L_{\mathrm{ASR}} + \beta L_{\mathrm{SWFC}} 
\label{equation:eq1}
\end{align}
where \( \alpha \) and \( \beta \) are  hyperparameter for each auxiliary task.

\subsection{Multi-task Learning Network}

The proposed MTL network includes a primary emotion recognition task and three auxiliary tasks. The emotion, gender, and speaker recognition branches each utilize a two-layer linear classification network, with each layer comprising LayerNorm, a ReLU activation function, and Dropout. Specifically, the ASR branch incorporates an LSTM network to enhance temporal modeling capabilities before the two-layer linear classification network. Inspired by \cite{b8} ,A unique feature of our approach is the dynamic fusion of  hidden layer representations as SSL features. In addition, we employ a co-attention mechanism to interactively fuse the outputs of the first linear layer from each auxiliary task with the corresponding features from the emotion recognition task. This fusion process enables the emotion recognition task to be effectively guided by auxiliary task information, with the final predictions generated through a linear classifier.

\subsection{Feature Fusion with Co-attention}

The co-attention module begins by mapping the features from each branch to the same dimension using a linear projection. Then it applies a dot-product attention mechanism to compute the interaction weights between the emotional features and each auxiliary task feature. Specifically, separate attention score matrices are calculated between the emotional features and the speaker, gender, and speech content features. These auxiliary features are weighted and aggregated through Softmax normalization. Finally, the dynamically fused multi-task representations are combined with the original emotional features, resulting in enhanced emotion representations that incorporate complementary cross-task information.

\subsection{SWFC Loss}

This study leaves behind a novel loss function, SWFC, specifically designed to mitigate class imbalance in classification tasks. SWFC builds upon the traditional Focal Contrastive Loss by introducing a dual weight adjustment mechanism that dynamically modifies both the focusing parameter and sample weight. This enables the simultaneous optimization of hard sample mining and class balance within a contrastive learning framework. The proposed method promotes the clustering of similar samples while enhancing the separation of dissimilar ones by constructing positive and negative pairs in the feature space. The focusing parameter directs the model's attention to hard-to-classify samples, while the sample weight modulates the penalty for underrepresented classes, thereby reducing model bias caused by long-tailed distributions. The SWFC loss is defined as follows:

\begin{align}
L_{\mathrm{SWFC}} &= -\frac{1}{N} \sum_{i=1}^{N} w_{i} \cdot \left( \sum_{j \neq i} \frac{\exp \left( \frac{s_{ij}}{\tau} \right)}{\sum_{k \neq i} \exp \left( \frac{s_{ik}}{\tau} \right)} \right. \nonumber \\
&\quad \cdot \left. \left( 1 - \frac{\exp \left( \frac{s_{ij}}{\tau} \right)}{\sum_{k \neq i} \exp \left( \frac{s_{ik}}{\tau} \right)} \right)^{\gamma} \right)
\label{equation:eq2}
\end{align}
where \( w_i \) is the category weight of sample \( i \), \( s_{ij} \) is the dot product similarity between sample \( i \) and sample \( j \), \( \tau \) is the sample parameter controlling the intensity of the penalty, and \( \gamma \) is the focusing parameter.

\section{EXPERIMENTS AND RESULTS}
\subsection{Datasets}
The experiments in this study employ data from the MSP-Podcast Challenge \cite{b18}, a large, naturalistic speech emotion corpus consisting of speech segments collected from an audio-sharing website. The dataset is divided into training, development, and test sets, with 66992, 25258, and 3200 utterances, respectively. These utterances represent a range of emotions, including anger (A), happiness (H), sadness (S), fear (F), surprise (U), contempt (C), disgust (D), and neutral (N). To improve the precision of transcription in the original dataset, we retranscribe the text using the Whisper-large v3 model \footnote{https://huggingface.co/openai/whisper-large-v3}.

To assess the generalizability of the model, we also incorporate additional test data from the Emotiw2018 \footnote{https://sites.google.com/view/Emotiw2018} and MEIJU2025 Track1 English datasets \footnote{https://ai-s2-lab.github.io/MEIJU2025-website/}. Model performance is evaluated using three metrics: F1-Macro, F1-Micro, and Accuracy. These metrics offer a comprehensive assessment of the model's performance in SER.

\subsection{Implementation Details}

This study is based on the PyTorch \cite{b19}, with the MSP-Podcast Challenge serving as the baseline for comparative experiments. We utilize WavLM-large with a feature dimension of 1024 for our experiments. The experiments are conducted on an NVIDIA A100 40GB GPU cluster, with a batch size of 16. The model is trained for a total of 60 epochs, using an initial learning rate of 1e-5 and the Adam optimizer \cite{b20} for parameter updates.

\subsection{Results and Analysis}
\subsubsection{ Comparasion with Baseline}
As shown in Table 1, the results of the comparative analysis reveal significant improvements in performance metrics for the proposed approach compared to the baseline model on all three test sets. The baseline model relies on single-task fine-tuning, where the SSL model is only fine-tuned for emotion recognition. In contrast, the MTL approach develops in this study achieves notable improvements across all test sets: F1 Macro increases by 5.66\%, 3.20\%, and 5.14\%, while Accuracy and F1 Micro improve by 3.22\%, 3.67\%, and 4.57\%, respectively. Importantly, our method consistently outperforms the baseline by over 3\% on both core evaluation metrics—F1 Macro and Accuracy—demonstrating the effectiveness of the MTL framework in enhancing SER performance.

\subsubsection{ Ablation Study on Highlighted Modules}
This section presents systematic ablation experiments to assess the impact of each innovative component. As shown in Table 2, three comparison groups are created: (1) a pure MTL framework, (2) a MTL network without the co-attention mechanism, and (3) a baseline model that only implements the SWFC loss. Due to submission limits, evaluations are conducted on the Emotiw2018 and MEIJU2025 test sets, with the top and bottom rows of data for each experiment corresponding to these datasets.

The quantitative results demonstrates the positive contributions of each component. Compared to the baseline, the SWFC loss alone improves F1-Macro by 0.81\%–1.00\% , highlighting its effectiveness in mitigating category imbalance. The full multi-task framework, however, achieves a substantial performance improvement of 3.01\%–4.86\%, highlighting the effectiveness of cross-task knowledge transfer. Notably, the addition of the co-attention mechanism boosts F1-Macro by 0.33\%–2.81\%, demonstrating the importance of dynamically integrating speaker characteristics, gender attributes, and ASR information as a key driver of performance improvement. 
\begin{table}
\centering
\caption{Comparison between Baseline and Proposed models on three test sets.}
\label{tab1}
\setlength{\tabcolsep}{4pt} % 平衡列间距
\renewcommand{\arraystretch}{1.1} % 优化行高
\begin{tabular}{@{}p{2cm} c c c c@{}} 
\toprule
\textbf{Test Set} & \textbf{Models} & \textbf{F1 Macro} & \textbf{F1 Micro} & \textbf{Accuracy} \\ 
\midrule
\multirow{2}{*}{MSP-podcast} 
 & Baseline  & 29.83  & 33.84 & 33.84 \\ 
 & Proposed  & \textbf{35.49} & \textbf{36.75} & \textbf{36.75} \\ 
\midrule % 完整横线分割
 
\multirow{2}{*}{Emotiw2018} 
 & Baseline  & 26.21  & 38.75 & 38.75 \\ 
 & Proposed  & \textbf{29.41} & \textbf{41.42} & \textbf{41.42} \\ 
\midrule % 完整横线分割

\multirow{2}{*}{MEIJU2025} 
 & Baseline  & 16.79  & 23.60 & 23.60 \\ 
 & Proposed  & \textbf{21.93} & \textbf{28.17} & \textbf{28.17} \\ 
\bottomrule
\end{tabular}
\end{table}

\begin{table}
\centering
\caption{ Ablation study of MTL,Co-Attention,and SWFC Loss, where Emotiw2018 is above and MEIJU2025 is below.}
\label{tab2}
\begin{tabular}{@{}lcccccc@{}}
\toprule
\textbf{Approach} & \textbf{F1 Macro} & \textbf{F1 Micro} & \textbf{Accuracy} \\ 
\midrule
\multirow{2}{*}{Baseline} & 26.21 & 38.75 & 38.75 \\ 
                               & 16.79 & 23.60 & 23.60 \\
\midrule
\multirow{2}{*}{Ours w/o MTL} & 27.02 & 38.84 & 38.84 \\ 
                               & 17.79 & 23.38 & 23.38 \\ 
\midrule
\multirow{2}{*}{Ours w/o Co-attention} & 28.89 & 40.41 & 40.41 \\ 
                                   & 20.60 & 26.36 & 26.36 \\ 
\midrule
\multirow{2}{*}{Ours w/o SWFC Loss} & 29.22 & 41.06 & 41.06 \\ 
                               & 21.65 & 27.66 & 27.66 \\ 
\midrule
\multirow{2}{*}{Ours} & \textbf{29.41} & \textbf{41.42} & \textbf{41.42} \\ 
                       & \textbf{21.93} & \textbf{28.17} & \textbf{28.17} \\ 
\bottomrule
\end{tabular}
\end{table}

\subsubsection{Analysis with Different Auxiliary Task Combination}
In this section, we aim to explore how cross-task correlations influence emotion recognition. To achieve this, we conduct an ablation experiment involving task combinations, as the official test sets lacked any labeling. As a result, our experiments are limited to the Emotiw2018 and MEIJU2025 test sets. Notably, due to training on the challenge dataset, Speaker IDs are unique and do not overlap across the dataset. So Speaker Verification achieved an accuracy of 0 on both test sets, and these results are not presented here.

\begin{table*}[htbp]
\centering
\caption{Task Combination Analysis on Emotiw2018 and MEIJU2025 Datasets.}
\label{tab3}
\begin{tabular}{@{}l c c c c c c c@{}}
\toprule
\multirow{2}{*}{\textbf{Test Set}} & \multicolumn{4}{c}{\textbf{Task}} & \multicolumn{1}{c}{\textbf{SER}} & \multicolumn{1}{c}{\textbf{ASR}} & \multicolumn{1}{c}{\textbf{Gender}} \\
\cmidrule(lr){2-5} \cmidrule(lr){6-6} \cmidrule(lr){7-7} \cmidrule(lr){8-8}
\cmidrule(lr){6-8} % 让第二行的横线连在一起
 & SER & ASR & Gender & Speaker & Accuracy & WER & Accuracy \\ 
\midrule
\multirow{4}{*}{Emotiw2018}
& $\checkmark$ & $\times$ & $\times$ & $\times$ & 38.75 & -- & -- \\
& $\checkmark$ & $\checkmark$ & $\times$ & $\times$ & 40.28 & \textbf{30.68} & -- \\
& $\checkmark$ & $\times$ & $\checkmark$ & $\times$ & 36.42 & -- & 89.79 \\
& $\checkmark$ & $\times$ & $\times$ & $\checkmark$ & 36.84 & -- & -- \\
& $\checkmark$ & $\checkmark$ & $\checkmark$ & $\checkmark$ & \textbf{41.42} & 32.14 & \textbf{92.34} \\
\midrule
\multirow{4}{*}{MEIJU2025}
& $\checkmark$ & $\times$ & $\times$ & $\times$ & 23.60 & -- & -- \\
& $\checkmark$ & $\checkmark$ & $\times$ & $\times$ & 26.66 &\textbf{23.57} & -- \\
& $\checkmark$ & $\times$ & $\checkmark$ & $\times$ & 23.38 & -- & 96.28 \\
& $\checkmark$ & $\times$ & $\times$ & $\checkmark$ & 23.18 & -- & -- \\
& $\checkmark$ & $\checkmark$ & $\checkmark$ & $\checkmark$ & \textbf{28.17} & 25.56 & \textbf{98.80} \\
\bottomrule
\end{tabular}
\end{table*}

As shown in Table 3, experiments on the Emotiw2018 and MEIJU2025 cross-domain test sets reveal a decrease in emotion recognition accuracy compared to the baseline when either Speaker Verification or Gender Classification is used as a secondary task. This suggests that these tasks may not have a strong association with emotion recognition. However, when all three auxiliary tasks—ASR, Gender Classification, and Speaker Verification are combined, the model achieves optimal performance. Despite the suboptimal results of the gender and speaker classification tasks when considered individually, the ASR task, supported by its primary auxiliary function, provided acoustic features that significantly enhanced the model's robustness.

\subsubsection{Last HiddenState vs. FusionState}
As shown in Table 4, the effectiveness of the learnable weighted fusion mechanism is evaluated. The experimental setup includes (1) a static single-layer feature extraction strategy, using only the last-layer hidden state of WavLM-Large, and (2) a dynamic hierarchical feature fusion strategy, employing the parametric weighted fusion module introduced in this paper.

The evaluation results demonstrate that the learnable weighting mechanism improves F1-Macro by 1.67\%, 1.91\%, and 2.18\% across the three test sets, respectively. These results highlight the complementary role of paralinguistic information across different transformer layers. The weighting module automatically assigns contribution weights to each hidden layer through end-to-end optimization, effectively addressing the issue of phonetic paralinguistic information dilution caused by deep semantic features. This mechanism provides an optimized aggregation path for hierarchical acoustic representations, significantly enhancing the model's ability to capture emotionally relevant acoustic cues.

\begin{table}
\centering
\caption{Comparison results on three test sets using only the last hidden state features and the features obtained using a learnable weighted fusion strategy. Last denotes the last hidden state and Learnable denotes the weighted fusion features.}
\label{tab4}
\small % 缩小字体
\setlength{\tabcolsep}{2pt} % 压缩列间距（默认6pt）
\renewcommand{\arraystretch}{1.00} % 压缩行高（默认1.0）
\begin{tabular}{@{}c c c c c@{}} 
\toprule
\textbf{Test Set} & \textbf{Features Type} & \textbf{F1 Macro} & \textbf{F1 Micro} & \textbf{Accuracy} \\ 
\midrule
\multirow{2}{*}{MSP-podcast} 
 & Last      & 33.82 & 36.00 & 36.00 \\ 
 & Learnable & \textbf{35.49} & \textbf{36.75} & \textbf{36.75} \\ 
\midrule
\multirow{2}{*}{Emotiw2018} 
 & Last      & 27.50 & 39.02 & 39.02 \\ 
 & Learnable & \textbf{29.41} & \textbf{41.42} & \textbf{41.42} \\ 
\midrule
\multirow{2}{*}{MEIJU2025} 
 & Last      & 19.75 & 26.63 & 26.63 \\ 
 & Learnable & \textbf{21.93} & \textbf{28.17} & \textbf{28.17} \\ 
\bottomrule
\end{tabular}
\end{table}

\section{Conclusion}
This study presents an innovative solution for the Interspeech2025 Natural Scene Speech Emotion Recognition Challenge. The proposed MTL framework combines emotion recognition, gender classification, speaker verification, and speech recognition. It dynamically models the interactions between emotional and auxiliary task features through a collaborative attention mechanism, enabling context-aware feature fusion. To address challenges in recognizing minority class samples and distinguishing semantically similar samples, we introduce a sample-weighted focus-contrast loss function. This function strengthens the learning of hard-to-distinguish and minority class samples by dynamically adjusting sample weights, effectively mitigating issues related to category imbalance and semantic confusion. Experimental results demonstrate that the proposed approach significantly enhances emotion recognition performance. In future work, we plan to explore the interrelationships between additional speech paralinguistic features and emotion, and further improve emotion recognition through novel cross-feature fusion techniques.

\section{Acknowledgements}
This work is supported by the National Key Research and Development Program of China (No.2022YFF0608504).

\bibliographystyle{IEEEtran}

\end{document}